\documentclass[a4paper,fleqn]{cas-dc}

\usepackage[authoryear,longnamesfirst]{natbib}

\def\tsc#1{\csdef{#1}{\textsc{\lowercase{#1}}\xspace}}
\tsc{WGM}
\tsc{QE}
\tsc{EP}
\tsc{PMS}
\tsc{BEC}
\tsc{DE}
\begin{document}
\let\WriteBookmarks\relax
\def\floatpagepagefraction{1}
\def\textpagefraction{.001}

% Short title
\shorttitle{Landauer principle and cosmology}

% Short author
\shortauthors{Oem Trivedi}

% Main title of the paper
\title [mode = title]{Landauer's principle and information at the cosmological horizon}

\author[1]{Oem Trivedi}[]
\cormark[1]

\ead{oem.t@ahduni.edu.in}

\affiliation[1]{organization={International Centre for Space and Cosmology},
    addressline={Navrangpura}, 
    city={Ahmedabad},
    postcode={380009},
    country={India}}
    
\cortext[cor1]{Corresponding author}

% Here goes the abstract
\begin{abstract}
We show that the information loss at the cosmological apparent horizon in an expanding universe does not obey Landauer's principle for efficient information erasure. This is in contrast to what happens for black holes, where the principle is upheld exactly. We discuss about the implications this result provides for the differences between information loss at black holes and that at the cosmological apparent horizon, which we term the "Cosmological Information Paradox". We also discuss how this result may point towards incompleteness of quasi-local descriptions of energy and could also point towards different thermodynamic settings in gravitational systems and how information thermodynamics as we know it may have limits to its applicability.

\end{abstract}

% Keywords
% Each keyword is seperated by \sep

\maketitle

\section{Introduction}
The intersection of information theory and thermodynamics has long provided profound insights into the nature of physical systems. A cornerstone of this intersection is the Landauer Principle \cite{landauer1961irreversibility}, which posits that the erasure of one bit of information in a computational process is accompanied by a dissipation of energy at least \( k_B T \ln 2 \), where \( k_B \) is the Boltzmann constant and \( T \) is the temperature of the system. This principle fundamentally links information processing with thermodynamic irreversibility, indicating that information loss is inherently accompanied by entropy increase and energy dissipation \cite{norton2011waiting,bennett1982thermodynamics}. There has been quite some work towards the theoretical foundations and the experimental backings of the Landauer principle in recent years \cite{galisultanov2017capacitive,berut2012experimental,plenio2001physics,sagawa2008second,jun2014high}. The Landauer limit, in essence, sets a fundamental thermodynamic constraint on the efficiency of computation, asserting that no computation can be entirely free of energy dissipation. This principle is pivotal in understanding the physical limitations of classical and quantum computation, as it establishes a direct link between information loss and heat generation.
\\
\\
Information loss has been a topic of profound discussion in the black hole literature \cite{Raju:2020smc}, and so it may seem natural for one to think that the Landauer principle would be the subject of various discussions in that context. However, there have only been a few works which have addressed the Landauer limit with regards to gravitational physics. There is a review of the Landauer principle in gravitational physics in \cite{Herrera:2020dyh}, which does not directly address black hole thermodynamics in any way. Most of 
 the existing works primarily focus on information loss in the exterior during black hole accretion with Fuchs \cite{fuchs1992landauer} being the first towards this area, attempting to leverage Landauer's principle to validate the Bekenstein-Hawking entropy formula's constant as one-quarter of the event horizon area but this work doesn't address Hawking radiation. A refined approach was presented in \cite{song2008information} by Song and Winstanley, where they considered the black hole's exterior quantum system in thermal equilibrium with emitted Hawking radiation and their work demonstrated how a limited amount of entropy, under Landauer constraints, can fall into the black hole while maintaining compatibility with the generalized second law of black hole thermodynamics.Another interesting study was done  by Kim et.al. \cite{kim2010black}, where they considered the black hole as a quantum eraser. A compilation of various results concerning black hole entropy and the Landauer Principle using an entropic framework is provided in    
 \cite{denis2023entropy}. Menin in \cite{menin2023black} sought to look connections between the Bekenstein bound \cite{bekenstein1981universal} and the Landauer principle. Recently, Cortes and Liddle \cite{cortes2024hawking} were able to prove that the Landauer limit is exactly saturated in Hawking radiation of black holes.  
\\
\\
While we have seen a few works( one could argue quite little) on the black hole connections of the Landauer principle, the cosmological connections have not been explored to any appreciable extent. One work by Lee et.al.\cite{lee2007quantum} explores a holographic dark energy model considering Landauer constraints. Gough also considered an approach to holographic dark energy considering the Landauer principle, although from a slightly different perspective in \cite{gough2011holographic}. In this work we would like to now consider the direct connections of the Landauer principle and information loss in cosmology. This work explores the application of the Landauer Principle to the cosmological apparent horizon and aims to explore whether the information loss associated with the apparent horizon, characterized by the Gibbons-Hawking temperature, adheres to the Landauer limit. The implications of the results here can be far reaching.
\section{Landauer Limit on the Cosmological Horizon}
The Bekenstein-Hawking entropy formulation states that the entropy of a black hole is proportional to its surface area \cite{bekenstein1973black}. The formula for the entropy takes the form \begin{equation}\label{firstent}
    \frac{S}{k_{b}} = \frac{c^3 \pi r_{s}^2}{G \hbar}
\end{equation}
where $k_{b}$,c, G, $r_{s}$ and $\hbar$ refers to the Boltzmann constant, the speed of light, Newton's gravitational constant, the horizon radius and the reduced Planck's constant. One would note here that we are considering all physical constants and not taking on natural Planck units ($m_{p}=1$) for this work. On the basis of the entropy form, one can then study the evaporation of black holes through Hawking radiation \cite{hawking1974black}, showing that black holes radiate at the Hawking temperature. One can study the entropy of the universe instead of just black holes using the Bekenstein-Hawking formulation itself \cite{Simovic:2023dhg}. A way to go by doing that is to recognize the horizon as that of the cosmological apparent horizon, instead of the event horizon of the black hole. In that case one has \begin{equation}
    r_{s} = \frac{c}{H}
\end{equation}
Using which one can arrive at the entropy form of \eqref{firstent} as \begin{equation} \label{secondent}
    \frac{S}{k_{b}} = \frac{c^5 \pi }{G \hbar H^2}
\end{equation}
The equivalent of the Bekenstein-Hawking temperature for the cosmological event horizon is the Gibbons-Hawking temperature \cite{gibbons1977cosmological}. In the context of de Sitter space, this temperature is associated with the cosmological event horizon, similar to how the Hawking temperature is associated with the event horizon of a black hole. The Gibbons-Hawking temperature is given by \begin{equation} \label{ght}
    T_{GH} = \frac{\hbar H}{2\pi k_B}
\end{equation}
In a de Sitter universe, the cosmological horizon temperature reflects the thermal nature of the horizon due to quantum effects in curved spacetime. This temperature is important in understanding the thermodynamics of the universe's expansion, paralleling the thermodynamics of black hole horizons. One should also note that there can be another description for the Horizon temperature, which is the Cai-Kim temperature \cite{cai2009hawking}. While the Gibbons-Hawking temperature is associated with the cosmological event horizon in a universe with a positive cosmological constant, the Cai-Kim temperature is more broadly applied to the apparent horizon in various cosmological models, including those that may not have a cosmological constant or may have additional forms of matter and energy. The apparent horizon in these cases is a surface that dynamically responds to the evolution of the universe, rather than being fixed like the cosmological event horizon in de Sitter space.
However, in the standard de Sitter case for a cosmological constant universe, the Gibbons-Hawking temperature suffices and here, we would like to address the standard case and would like to leave the treatment for other dark energy models with Cai-Kim treatments open. 
From \eqref{firstent}, we can write \begin{equation}
    \frac{\Delta S}{S} = - \frac{2 \Delta H}{H}
\end{equation}
The change in H while losing one Boltzmann unit of entropy will then be given as \begin{equation}
    \Delta H = - \frac{H k_{b}}{2 S}
\end{equation}
Now, the conversion between the number of bits and entropy in Boltzmann units provides us with the ln 2 factor here which gives us \begin{equation} 
    \Delta H = - \frac{H k_{b} \ln{2}}{2 S}
\end{equation}
Using \eqref{firstent}, we can write \begin{equation} \label{delh}
    \Delta H = - \frac{G \hbar H^3 \ln{2}}{2 \pi c^5}
\end{equation}
Now in order to address the energy of the apparent horizon, one has to think carefully. In the framework of general relativity, defining energy is inherently complex due to the curvature of spacetime. Unlike classical physics, where energy can be easily isolated and quantified, the dynamic nature of spacetime under gravity makes it difficult to pinpoint a clear global energy definition. Quasi-local energy addresses this challenge by focusing on the energy contained within a finite, bounded region of spacetime. This concept is crucial for understanding how energy is distributed and how gravitational fields influence it, providing insights into the behavior of intricate gravitational systems like black holes, cosmological horizons, and gravitational waves. Quasi-local energy definitions serve as a bridge between local properties (like curvature at a point) and global properties (like total mass or energy of the universe), offering a more practical approach to studying the energetic aspects of spacetime. There has been long discussions about proper quasi-local definitions of energy which are valid in a bounded region of space, and a for a good review of this one can refer to \cite{szabados2009quasi}. A useful definition in this regard is that of Misner and Sharp \cite{misner1964relativistic}. The Misner-Sharp energy \(E\) is defined in a spherically symmetric spacetime and possesses several useful properties, making it a reasonable approach for studying quasi-local energy. In the Newtonian limit of a perfect fluid, \(E\) corresponds to the Newtonian mass at leading order and represents the Newtonian kinetic and potential energy at the next order. For test particles, the related Hajicek energy is conserved and behaves as expected in both Newtonian and special-relativistic limits. In the small-sphere limit, the leading term in \(E\) is the product of the volume and the energy density of the matter. In vacuum, the Misner-Sharp energy reduces to the Schwarzschild energy. At null and spatial infinity, \(E\) reduces to the Bondi-Sachs and ADM energies, respectively. Notably, the conserved Kodama current generates the conserved charge \(E\). For these reasons, the Misner Sharp energy has been seen as quite a valid form for quasi-local studies, and has been very useful for studying FLRW cosmology as well in a variety of theories \cite{sanchez2023thermodynamics,kong2022pv,moradpour2016thermodynamic,hu2015misner,zhang2021thermodynamics,Maeda_2008,Cai_2009}. For a spherically symmetric metric, it is given by \begin{equation}
    E = \frac{c^4}{2G} R \left( 1 - g^{\mu\nu} \partial_\mu R \partial_\nu R \right)
\end{equation}
At the apparent horizon, the term \( g^{\mu\nu} \partial_\mu R \partial_\nu R \) vanishes (since the apparent horizon is a null surface) and therefore, the Misner-Sharp energy simplifies to \begin{equation}
     E = \frac{c^4 R}{2G}
\end{equation}
Plugging the radius of the apparent horizon $R=r_{s}$ into the Misner-Sharp energy formula, we get \begin{equation}
    E = \frac{c^4 r_{s}}{2G} = \frac{c^4}{2G} \frac{c}{H} = \frac{c^5}{2GH}
\end{equation}
Using this, we can then write \begin{equation}
    \Delta E = - \frac{c^5}{2 G H^2} \Delta H 
\end{equation}
We can utilize \eqref{delh} now, to then write \begin{equation} \label{main1}
    \Delta E = -\frac{c^5}{2 G H^2} \left(-  \frac{G \hbar H^3 \ln{2}}{2 \pi c^5} \right) = \frac{1}{2} \frac{\hbar H}{2 \pi } \ln{2}
\end{equation}
which allows us to write using \eqref{ght} \begin{equation} \label{main2}
    \Delta E = \frac{1}{2} k_{b} T_{GH} \ln{2}
\end{equation}
we see that this is exactly half of the minimum constraint applied by the Landauer principle, for radiation at the Gibbons-Hawking temperature. From this we can conclude that the Landauer limit is \textit{not satisfied} in the cosmological context, which is contrary to the Black hole case where it is indeed satisfied \cite{cortes2024hawking}.  
\section{Conclusions}
In this work, we have explored the implications of the Landauer Principle in the context of the cosmological apparent horizon, drawing a parallel with the well-studied case of black hole thermodynamics. Our primary focus was on the energy associated with the apparent horizon and its compliance, or lack thereof, with the Landauer limit. Note that the Landauer limit is derived fundamentally from the first law of thermodynamics itself \cite{landauer1961irreversibility} and so has a direct correspondence in its essence with basic thermodynamic principles. The main results of our work as, seen in \eqref{main1}-\eqref{main2}, can lead one to several interesting conclusions and implications at the intersections of various sub-fields. We highlight these possible conclusions as follows : 
\begin{itemize}
    \item Incompleteness of quasi-local descriptions ? : The energy E in \eqref{main1}-\eqref{main2} considered in our work is the quasi-local energy associated with the apparent horizon, defined through the Misner-Sharp mass formula. This energy represents the total gravitational energy within a finite, bounded region of spacetime, encompassing both the kinetic and potential energy contributions and in cosmological terms, this energy provides a measure of the total mass-energy content within the horizon, offering a quasi-local description of the universe's energetic state. Our results suggest that the quasi-local descriptions of energy, as encapsulated by the Misner-Sharp energy, may be incomplete or insufficient in capturing the full thermodynamic behavior of the universe. The discrepancy with the Landauer limit implies that the energy dissipation associated with the cosmological apparent horizon does not operate at the most fundamental level of efficiency expected from information theory principles. This raises the possibility that our current theoretical frameworks, particularly those relying on quasi-local energy concepts, might not fully account for all aspects of gravitational thermodynamics on cosmological scales.
    \item Fundamental differences between Black hole and cosmological horizon thermodynamics? Black hole vs Cosmological information paradox ? : It has been a prevalent feature of black hole thermodynamics that a lot of its core ideas can be implemented in cosmology almost seamlessly, as highlighted by various theories of both the late and the early universe which have been derived from thermodynamic principles. However, here we seeing hints that perhaps there are some fundamental differences in both regimes which need to be explored more. The adherence to the Landauer limit in black holes implies that even if information is lost, it is done so in a maximally efficient way \cite{cortes2024hawking} while in contrast, our work shows that the cosmological apparent horizon does not satisfy the Landauer limit. The energy change associated with one bit information loss for the cosmological horizon is only half of the minimum required by the Landauer Principle, suggesting that the universe as a whole is not as efficient in information processing as black holes. This discrepancy raises questions about the nature of information loss on a cosmic scale and it may suggest the existence of what might be termed a "cosmological information paradox," analogous to the black hole information paradox. While black holes might lose or transform information efficiently, the universe's apparent inefficiency suggests that the processes governing information loss or transformation on cosmological scales are fundamentally different. So it could be possible that the ways to look for retrieval of information lost to the cosmological horizon can be very different from the efforts put towards retrieving information loss at black holes and suggests that significant work needs to be undertaken in that direction. 
    \item Different thermodynamic settings in gravitational physics? : The divergence between the black hole case, where the Landauer limit is satisfied, and the cosmological case, where it is not, may also hint at a deeper underlying difference in thermodynamic settings for different gravitational systems. While black holes seem to process and dissipate information in an optimally efficient manner, the universe when viewed as a whole appears to operate under different constraints or principles. This discrepancy may indicate that fundamental gravitational physics could vary across different scales or regimes, with distinct thermodynamic behaviors emerging in cosmological settings compared to localized systems like black holes. This could also suggest that the principles governing energy dissipation and information loss may not be universally consistent across all gravitational systems, indicating that the current theoretical framework might be limited or incomplete when applied to the entire universe.
    \item Extended entropy formulations or non $\Lambda$ dark energy better suited? : Finally, one can also think of whether the results here would hold when one considers extensions of the Bekenstein-Hawking entropy \cite{tsallis2013black,kaniadakis2002statistical,renyi1961measures,Barrow:2020tzx,jalalzadeh2021prospecting,Jalalzadeh:2022rxx,_imdiker_2023,nojiri2022early,nojiri2023microscopic,nojiri2022modified,odintsov2023non,odintsov2024second,nojiri2024horizon,odintsov2024primordial,odintsov2023holographic}. It could be possible that extended gravity effects could readily provide one with scenarios where $ \Delta E \geq k_{b} T_{GH} \ln{2}$, given free parameters from the underlying theories etc. ( like in the case of Tsallis, Barrow, Kaniadakis and other formulations). Even going beyond the usual extended Bekenstein-Hawking relations, one can even consider the generalized Nojiri-Odintsov free parameter model as well. In the event that it may occur that the general entropies allow one to be consistent with the Landauer limit, it could suggest that the corrected Bekenstein-Hawking formulas hold better when dealing with the cosmological horizon than the standard formulation. In these cases, one would also possibly take into account the Cai-Kim temperature of the horizon instead of the Gibbons-Hawking one that we took here. Even beyond this, a fundamental assumption in our work as we noted before was that we were considering the scenario of a positive cosmological constant, which led to the applicability of the Gibbons-Hawking temperature in this regard. So it could also be an avenue of fruitful thought to now investigate if various non $\Lambda$ models of dark energy can then lead towards satisfying the Landauer limit which could hint towards them being better suited too.  
\end{itemize}
\section*{Acknowledgements}
The author would like to thank Sunny Vagnozzi, Sergei Odintsov and Tanmoy Paul for very important discussions.

	\bibliographystyle{cas-model2-names}
	\bibliography{cas-refs} 

\end{document}